\documentclass[submission,copyright]{eptcs}
\providecommand{\event}{LFMTP'26} % Name of the event you are submitting to

\usepackage{iftex}

\ifpdf
  \usepackage{underscore}         % Only needed if you use pdflatex.
  \usepackage[T1]{fontenc}        % Recommended with pdflatex
\else
  \usepackage{breakurl}           % Not needed if you use pdflatex only.
\fi
\usepackage[utf8]{inputenc}
\usepackage{graphicx}
\usepackage{hyperref}
\usepackage{color}
\usepackage{xspace}
\usepackage[inline]{enumitem}

\usepackage{stmaryrd}
\usepackage{amsmath}
\usepackage{amssymb}
\usepackage{cmll}
\usepackage{simplebnf}
\usepackage{bussproofs}

\newcommand{\ie}{i.e.,\xspace}

\DeclareMathOperator{\Seq}{Str}
\DeclareMathOperator{\States}{\Sigma}

\DeclareMathOperator{\Flatmap}{\mathbin{\gg\!=}}
\DeclareMathOperator{\Diag}{\triangle}
\DeclareMathOperator{\If}{if}
\DeclareMathOperator{\Owise}{otherwise}
\DeclareMathOperator{\lb}{\llbracket}
\DeclareMathOperator{\rb}{\rrbracket}
\DeclareMathOperator{\Id}{id}

\DeclareMathOperator{\SeqC}{;}
\DeclareMathOperator{\Par}{\|}
\DeclareMathOperator{\FSeqC}{\with\!;}
\DeclareMathOperator{\FPar}{\with\!|}

\DeclareMathOperator{\Mgu}{mgu}

\title{A Strategy Language for Controlled Proof Search}
\author{
Romain~Sidhoum
\qquad
Simon~Robillard
\qquad
David~Delahaye
\\
\institute{LIRMM, Univ. Montpellier, CNRS, Montpellier, France}
\email{Firstname.Lastname@lirmm.fr}}
\def\titlerunning{A Strategy Language for Controlled Proof Search}
\def\authorrunning{R. Sidhoum, S. Robillard, D. Delahaye}

\begin{document}
\maketitle

\begin{abstract}
This paper introduces the strategy language of Pgeon, a meta-prover with a clear separation between inference rules and proof search. We give the semantics of strategies as functions over proof states, and of the operators that are used to combine them, allowing for sequential composition, choice, repetition and interleaving of strategies.
This language is designed to handle the challenge of fair proof search in semi-decidable logics, where simple depth-first exploration of the proof space is not guaranteed to achieve completeness.
We showcase the expressiveness and effectiveness of the approach through case studies in first-order and modal logics.
%\keywords{Tableau Calculi \and Strategy Languages \and Logical Framework}
\end{abstract}

\section{Introduction}
In automated reasoning, the completeness of a logical calculus does not guarantee that a proof will be found.
In semi-decidable logics, some rules may generate infinitely many successors, and unfair exploration may indefinitely postpone relevant inferences, preventing the discovery of valid proofs.
The completeness of the calculus is sometimes referred to as \emph{static completeness}, while \emph{dynamic completeness} describes the property of a prover, \ie a proof search algorithm applied to a given calculus, to find a proof for any theorem~\cite{waldmann2020comprehensive}.

These issues are particularly visible in systems that separate inference rules from proof-search strategies.
While rules describe admissible transformations, strategies determine how they are applied during search, making the design of effective exploration policies non-trivial.

Pgeon~\cite{sidhoum2026pgeon} is a framework that separates logical rules from proof-search strategies and compiles them into executable provers. This separation highlights the impact of strategy design: the same set of rules may lead to different outcomes depending on how they are explored.

We introduce a formal strategy language for Pgeon, together with a compositional semantics that models proof search as the enumeration of proof states.
Strategies are interpreted as functions mapping states to (possibly infinite) streams of successors, capturing both deterministic and non-deterministic behavior.
The language provides combinators for sequential composition, choice, repetition, and fair interleaving of strategies.

We illustrate the approach through case studies in first-order and modal logics, showing how naive strategies may fail to terminate and how fair combinators enable systematic exploration of the search space.
More generally, the framework provides a basis for reasoning about proof-search procedures at an abstract level, independently of implementation details.

\paragraph{Related Work}
Our strategy language relates to work on compositional stream processing, monadic nondeterminism, and strategy languages.

Monoidal stream semantics \cite{di2022monoidal} models stream processors within a symmetric monoidal category, providing a principled account of compositionality where complex transformations are built from simpler ones. Similarly, functional reactive programming (FRP) \cite{elliott1997functional} treats stream transformers as compositional entities combined via structured interfaces. However, these approaches are typically deterministic, whereas our setting is inherently nondeterministic and allows multiple (possibly infinite) results, requiring explicit control over exploration.

The treatment of nondeterministic computations over streams is closely related to the LogicT framework \cite{kiselyov2005backtracking}, which provides a monadic encoding of backtracking with both biased and fair search. Sequential composition in our language corresponds to Kleisli composition, while our choice operators capture biased and fair exploration. In contrast to LogicT, which provides an operational abstraction, we reify computations as lazy streams and introduce an explicit algebra of strategy combinators.

Our work is also related to tactic languages such as $\mathcal{L}_{tac}$ \cite{delahaye2000tactic}, used in the prover Rocq, where tactics act as transformations over proof states with failure and backtracking. While the analogy is structural, tactic languages typically provide an operational notion of search. In our approach, the search space is represented explicitly as a stream, and exploration strategies are expressed compositionally.

Rewriting-based systems such as Maude \cite{clavel2007all} similarly model computation as nondeterministic transitions and provide strategy languages for controlling rule application. However, rewriting logic is fundamentally relational, whereas our approach is functional and stream-based, reifying the search space as a first-class object and enabling compositional reasoning about fairness and enumeration.

\section{Pgeon}
Pgeon~\cite{sidhoum2026pgeon} generates automated theorem provers from declarative specifications of logical calculi. It is intended to generate provers based on the tableaux method. The specification of a prover defines the syntax, inference rules, and a strategy that describes how rules are applied. From this specification, an executable prover is produced.

This separation allows different proof-search behaviors to be defined over the same calculus, as the rules specify admissible steps while strategies control their exploration.
In particular, rule applications are inherently non-deterministic, as they may match multiple formulas, branches, or terms, potentially generating infinitely many successors.

In Pgeon, this search policy is expressed by a strategy.
Strategies control the order of rule applications, the exploration of alternatives, and the use of backtracking, making them a central component of the generated prover.
The strategy language studied in Section \ref{sec:strategy} abstracts away from the concrete implementation of this mechanism.
We treat a rule application as a function from proof states to streams of successor states.
The stream represents the possible outcomes of applying the rule, ordered according to the intended exploration policy.
This view makes non-determinism explicit since all possible outcomes are enumerated rather than implicitly explored.
A deterministic rule returns a singleton stream, a failing rule returns the empty stream, and a non-deterministic rule returns a stream with several successors.
A rule may yield an infinite stream of successors.

This abstraction is useful because it allows strategy operators to be defined compositionally.
Sequential composition, choice, repetition, and fair interleaving can all be described as operations on streams of proof states.
The resulting semantics does not depend on the stack discipline or continuation representation used by particular implementations.

The need for such a semantics it becomes clear when considering fairness.
Suppose a strategy repeatedly applies a universally quantified rule before considering other instances.
Even if a contradiction can be found after a finite number of different instantiations, a depth-first strategy may keep generating redundant or irrelevant instances forever.
The problem is not that the calculus is incomplete, but that the strategy unfairly gives unbounded priority to one part of the search space.

For example, consider the clause set:
\[ 
\Gamma = \{ \; \forall x. \neg P(x) \lor \neg Q(x), \;
P(a), \;
P(b), \;
Q(b) \; \}
\]
A closing derivation requires selecting useful instances of the universal formula and combining them with the atomic facts already present in the branch.
However, repeated instantiation with $x \mapsto a$ produces an infinite expansion without progress.
The search then follows an infinite path although the contradiction is reachable in the proof space.
This phenomenon can be sketched as follows:
\begin{prooftree}
\rootAtTop
    \AxiomC{}
    \UnaryInfC{$\Gamma, \neg P(a)$}
        \AxiomC{}
        \UnaryInfC{$\Gamma, \neg Q(a), \neg P(a)$}

        \AxiomC{\ldots}
        \UnaryInfC{$\Gamma, \neg Q(a), \neg Q(a)$}
    \RightLabel{$\forall, \lor, x \mapsto a$}
    \BinaryInfC{$\Gamma, \neg Q(a)$}
\RightLabel{$\forall, \lor, x \mapsto a$}
\BinaryInfC{$\Gamma$}
\end{prooftree}

This motivates the fair strategy combinators introduced in the rest of the paper.
Rather than relying on depth-first exploration alone, the language provides operators that interleave the results of several strategies.
Interleaving ensures that if a successor state occurs at some finite position in one of the component searches, then it also occurs at some finite position in the combined search, so no branch that can produce a result is postponed indefinitely.
In this way, fair composition prevents one infinite branch from starving the exploration of another.

\section{Strategy Language}
\label{sec:strategy}
We formalize proof search as the transformation of proof states.
A proof state represents a partially constructed proof tree, whose open branches correspond to sets of formulas to be closed.
Applying an inference rule expands the proof tree, producing new proof states.

Let $\States$ be the set of proof states. In general, a proof state can be seen as a tree of formulas, whose branches are either open or closed. The precise structure of $\States$ is left abstract, as our framework is parametric in the underlying proof system.

Rather than modeling proof search as a relation on states, we adopt a view where a strategy enumerates all possible outcomes of its application.
Formally, we write $\Seq(\States)$ for the set of (finite or infinite) streams over $\States$, and define a strategy as a function:
$$ s : \States \rightarrow \Seq(\States) $$
Given a state $\sigma \in \States$, the stream $s(\sigma)$ represents all possible successor states obtained by applying $s$ to $\sigma$.
The empty stream denotes failure, a singleton stream denotes deterministic success, and longer streams represent multiple possible successor states (non-deterministic branching).

Each inference rule $r$ induces a primitive strategy:
$$\lb r \rb : \States \rightarrow \Seq(\States)$$
which returns all proof states obtained by applying $r$ for each admissible match of premises.
Depending on the rule, the matched premises may either be removed from the state, or preserved. In the latter case, it is often useful to apply the rule exhaustively to all applicable matches.
To this end, we introduce a derived operator $r!$, which applies $r$ to all admissible matches in a given state.
Its semantics is given by:
$$\lb r! \rb : \States \rightarrow \Seq(\States)$$
where $\lb r! \rb(t)$ returns the unique state obtained by exhaustively applying $r$ to all applicable premises in $t$, if at least one application is possible, and the empty stream otherwise.

\paragraph{Stream operators}
We assume the following operators on streams:
\begin{itemize}
    \item The stream operator $\Id$ such that $\Id(t) = \langle t \rangle $
    \item For a function $f : \States \rightarrow \Seq(\States)$ and a stream $X$, $X \Flatmap f$ is the stream obtained by applying $f$ to each element of $X$ and concatenating the results.
    \item Given a stream of streams $S = \langle S_0, S_1, S_2, \ldots \rangle$ with $S_i = \langle \sigma_{i,0}, \sigma_{i,1}, \ldots \rangle $, we define
%$ \Diag(S) = \langle \sigma_{i,j} | (i,j) \in \mathbb{N}^2 \text{ enumerated in increasing order of } i + j \text{ (breaking ties by increasing i), for all } (i, j) \text{ such that } \sigma_{i, j} \text{ exists.} \rangle $
$\Diag(S)$ the stream that contains every $\sigma_{i,j}$, ordered such that $\sigma_{i,j}$ appears before $\sigma_{i',j'}$ if
\begin{enumerate}[label=(\alph*)]
    \item $i + j < i' + j'$, or
    \item $i + j = i' + j'$ and $i < i'$.
\end{enumerate}
This construction ensures that every element $\sigma_{i, j}$ appears after finitely many elements in $\Diag(S)$. Thus, if a result appears at a finite depth in any component stream, it appears at a finite depth in the interleaved stream.
\end{itemize}

\paragraph{Strategy combinators}
Strategies are built from primitive rules ($r$) using the following grammar:
\begin{center}
\begin{bnf}
$s$ ::=
$r$ //
$r!$ //
$s \SeqC s$ // 
$s \Par s$ // 
$s \FSeqC s$ // 
$s \FPar s$ //
$s^*$
\end{bnf}
\end{center}

Let $s, s_1, s_2 : \States \rightarrow \Seq(\States)$. The semantics of the combinators is defined as follows.

\begin{itemize}
    \item Sequential composition:
$$ \lb s_1 \SeqC s_2 \rb(t) = \lb s_2 \rb \Flatmap \lb s_1 \rb (t) $$
This corresponds to applying $s_1$ to $t$, and then applying $s_2$ to each resulting state.

    \item Left-biased choice:
$$
\lb s_1 \Par s_2 \rb(t) =
\begin{cases}
\lb s_1 \rb (t) \quad \If \, \lb s_1 \rb (t) \neq \emptyset,\\
\lb s_2 \rb (t) \quad \Owise
\end{cases}
$$
This operator models deterministic choice with fallback.

\item Fair sequential composition:
$$ \lb s_1 \FSeqC s_2 \rb(t) = \Diag ( \lb s_1 \rb (t) \Flatmap \lb s_2 \rb ) $$ 
Unlike $\Par$, this operator interleaves the exploration of all intermediate results, ensuring that no branch is indefinitely delayed.
If $s_2(s_1(t))$ can produce a result in finite steps, then $s_1 \FSeqC s_2$ will eventually produce it.

\item Fair choice:
$$ \lb s_1 \FPar s_2 \rb(t) = \Diag (\langle \lb s_1 \rb (t), \lb s_2 \rb (t) \rangle) $$
This operator fairly interleaves the results of both strategies.

\item Repetition:
$$ s^*= \mu X. \Id \FSeqC (s \FSeqC X) $$
This defines the closure under repeated application of $s$, with fairness ensuring that all finite iteration depths are explored.
Unlike the usual Kleene star based on sequential composition, this definition uses fair sequential composition. As a result, the exploration of derivations is interleaved across all iteration depths, rather than proceeding depth-first.
The fixpoint is understood as the least fixpoint on strategies ordered by prefix inclusion of their output streams.

\end{itemize}
The operators $\SeqC$ and $\Par$ induce a depth-first exploration of the search space, as they process results sequentially and may indefinitely delay alternative branches.
The operators $\FSeqC$ and $\FPar$ use diagonalization to ensure a fair exploration, guaranteeing that every branch that produces results is eventually explored.

\section{Case Studies}
\subsection{First-Order Logic (LK)}
We illustrate the expressiveness of the strategy language on a tableau calculus for classical first-order logic (LK).
Inferences are classified into four families: non branching rules $\alpha$, branching rules $\beta$, universal rules $\gamma$, and existential rules $\delta$. The rules are given in Figure \ref{fig:lk}. This calculus uses a destructive closure rule $\odot$: the closing substitution is applied to all branches, potentially forbidding inferences that would have previously been possible. The closing rule is the only source of instantiation, \ie there is no $\gamma$ rule that can provide an instance $P(t)$ of some universally quantified formula.

\begin{figure}[ht]
\setlength{\belowcaptionskip}{-10pt}
\framebox[\textwidth][c]
{\parbox{\textwidth}
{\begin{center}
\begin{tabular}{c@{\hspace{0.5cm}}c}

\AxiomC{$P$}
\AxiomC{$\neg P$}
\RightLabel{$\odot$}
\BinaryInfC{$\odot$}
\DisplayProof&

\AxiomC{$\neg\neg P$}
\RightLabel{$\alpha\neg\neg$}
\UnaryInfC{$P$}
\DisplayProof\\\\

\AxiomC{$P \land Q$}
\RightLabel{$\alpha\land$}
\UnaryInfC{$P, Q$}
\DisplayProof&

\AxiomC{$\neg (P \lor Q)$}
\RightLabel{$\alpha\neg\lor$}
\UnaryInfC{$\neg P, \neg Q$}
\DisplayProof\\\\

\AxiomC{$\neg(P \Rightarrow Q)$}
\RightLabel{$\alpha\neg\Rightarrow$}
\UnaryInfC{$P, \neg Q$}
\DisplayProof&

\AxiomC{$P \lor Q$}
\RightLabel{$\beta\lor$}
\UnaryInfC{$P \hspace{1em} | \hspace{1em} Q$}
\DisplayProof\\\\

\AxiomC{$\neg (P \land Q)$}
\RightLabel{$\beta\neg\land$}
\UnaryInfC{$\neg P \hspace{1em} | \hspace{1em} \neg Q$}
\DisplayProof&

\AxiomC{$P \Rightarrow Q$}
\RightLabel{$\beta\Rightarrow$}
\UnaryInfC{$\neg P \hspace{1em} | \hspace{1em} Q$}
\DisplayProof\\\\

\AxiomC{$\exists x. P(x)$}
\RightLabel{$\delta\exists$}
\UnaryInfC{$P(f(X_1, \ldots, X_n))$}
\DisplayProof&

\AxiomC{$\neg \forall x. P(x)$}
\RightLabel{$\delta\neg\forall$}
\UnaryInfC{$\neg P(f(X_1, \ldots, X_n))$}
\DisplayProof\\\\

\AxiomC{$\forall x. P(x)$}
\RightLabel{$\gamma \forall$}
\UnaryInfC{$P(X)$}
\DisplayProof&

\AxiomC{$\neg \exists x. P(x)$}
\RightLabel{$\gamma \neg \exists$}
\UnaryInfC{$P(X)$}
\DisplayProof\\\\

\multicolumn{2}{c}{
\AxiomC{$P$}
\AxiomC{$\neg Q$}
\RightLabel{$\odot$}
\BinaryInfC{$\odot$, apply $\sigma$ to the tree if $P$ and $Q$ are unifiable and $\sigma = \Mgu(P, Q)$ }
\DisplayProof
}

\end{tabular}
\end{center}}}
\caption{LK Rules for First-Order Classical Logic}
\label{fig:lk}
\end{figure}

A common heuristic is to separate rule applications into phases: First, perform all applications of $\alpha$ and $\delta$ rules, then applications of $\beta$ rules. Interleave this process with instantiations of universal formulas. A naive strategy would consist in saturating the branch and repeatedly applying $\gamma$: 
$(\alpha \Par \delta \Par \beta \Par \gamma)^*$.

The main difficulty lies in the interaction between closure and the rest of the proof search.
Failing to consider certain closure choices may delay the discovery of useful instantiations. An effective strategy must therefore ensure that all relevant closure-induced substitutions are eventually explored.

To address this issue, we separate closure from the rest of the search and combine them using fair composition:
$$(\odot^* \FSeqC (\alpha \Par \delta \Par \beta \Par \gamma!)^*)^*$$
The sub-strategy $\odot^*$ enumerates all finite streams of closure applications, thereby exploring different possible instantiations induced by unification.
The sub-strategy $(\alpha \Par \delta \Par \beta \Par \gamma!)^*$ performs saturation using the remaining rules.
Because $\Par$ is left-biased, it can be used to prioritize certain rules.
The fair composition operator $\FSeqC$ interleaves these two processes, ensuring that closure attempts are distributed across all depths of the search.
As a result, the strategy avoids committing prematurely to a particular instantiation: if a closing derivation exists, it will eventually be explored.

\subsection{Modal Logic (S4)}
We now consider a second case study based on propositional modal logic S4 \cite{gore1999tableau}.
We consider a standard classification of tableau rules for propositional S4. In addition to the usual proposition rules ($\alpha$ and $\beta$), we use the three modal rules described in Figure \ref{fig:s4}.

\begin{figure}[ht]
\begin{minipage}{\textwidth}
\setlength{\belowcaptionskip}{-10pt}
\framebox[\textwidth][c]
{\parbox{\textwidth}
{\begin{center}
\begin{tabular}{c@{\hspace{0.5cm}}c@{\hspace{0.5cm}}c}

\AxiomC{$P$}
\RightLabel{$\theta$\footnote{This rule operates on a single branch. In pgeon, the tableau is represented as a forest of branches, enabling local transformation.}}
\UnaryInfC{}
\DisplayProof&

\AxiomC{$\Box P$}
\RightLabel{$T$}
\UnaryInfC{$\Box P; P$}
\DisplayProof&

\AxiomC{$\Box X; \neg \Box P$}
\RightLabel{$S4$\footnote{This rule is non-local, as it requires access to the entire branch, unlike the other rules which are applied to individual formulas.}}
\UnaryInfC{$\Box X; \neg P$}
\DisplayProof\\\\

\end{tabular}
\end{center}}}
\end{minipage}
\caption{S4 Rules for Propositional Modal Logic}
\label{fig:s4}
\end{figure}

The main source of non-determinism is the rule $\theta$, that allows arbitrary removal of formulas from the branch.
$\theta$ is necessary to expose formulas to the modal rule $S4$ by removing blocking formulas so that $S4$ becomes applicable, but it may also discard formulas that are needed later to close the branch.
An uncontrolled use of $\theta$ may therefore lead to dead ends in the search space.

The key observation is that $\theta$ should not be used arbitrarily, but only in order to enable applications of $S4$.
We capture this idea by grouping $\theta$ and $S4$ into a combined strategy:
$\theta^* \FSeqC S4$.
We combine this construction with the other rules to obtain the following strategy:
$$ ((\alpha \Par \beta \Par T)^* \SeqC (\theta^* \FSeqC S4))^* $$

At each iteration, the strategy first saturates using $\alpha$, $\beta$, and $T$, which are terminating using loop checking.
If no such rule is applicable, it applies the combined strategy which explores all ways of applying $S4$ with the removal of some formulas.
The use of fair composition ensures that different choices of formulas to discard are explored without starvation. If a successful application of $S4$ requires removing a specific subset of formulas, this configuration will eventually be reached. If closing the tableau additionally requires not discarding a specific subset of formulas, this configuration will also be reached.

\section{Conclusion}
This work introduced a formal strategy language for controlling proof search in tableau-based automated theorem proving, as implemented in Pgeon.
By modeling strategies as functions from proof states to streams of successor states, the framework makes non-determinism explicit and enables a compositional semantics for combining search behaviors.
The language provides both biased and fair combinators, with fairness achieved through interleaving constructions that prevent the starvation of branches in potentially infinite search spaces.
Through case studies in first-order and modal logics, we demonstrated how naive depth-first strategies may fail to find proofs despite the completeness of the underlying calculus, and support dynamic completeness by enabling systematic exploration of the search space.
Overall, this approach offers a principled and modular foundation for specifying, analyzing, and reasoning about proof-search procedures independently of implementation details, while addressing fundamental challenges inherent to semi-decidable logics.

\bibliographystyle{eptcs}

\bibliography{bibliography}

\end{document}